\documentstyle[12pt]{article}

\newlength{\dinwidth}
\newlength{\dinmargin}
\def\half{{\textstyle{1\over 2}}}
\def\bm#1{{\bf #1}}
\def\to{\rightarrow}
\def\ee{e^+e^-}
\def\ww{W^+W^-}
\def\fg{f_6^{\gamma}}
\def\fz{f_6^{Z}}
\def\cM{{\cal M}}
\def\MeV{{\rm MeV}}
\def\GeV{{\rm GeV}}
\def\GeVc2{{\rm GeV}/c^2}

\def\lapproxeq{\lower .7ex\hbox{$\;\stackrel{\textstyle
<}{\sim}\;$}}
\def\gapproxeq{\lower .7ex\hbox{$\;\stackrel{\textstyle
>}{\sim}\;$}}

\begin{document}
\titlepage
\begin{flushright}
DTP/96/54\\
hep-ph/9607420\\
July 1996 \\
\end{flushright}

\begin{center}
\vspace*{2cm}
{\Large \bf Constraining a $CP$-violating $WWV$ coupling from\\[2mm]
the $\ww$ threshold cross section at LEP2}\\
\vspace*{1cm}
V.\ C.\ Spanos$^a$ and W.\ J.\ Stirling$^{a,b}$ \\

\vspace*{0.5cm}
$^a \; $ {\it Department of Physics, University of Durham,
Durham, DH1 3LE }\\

$^b \; $ {\it Department of Mathematical Sciences, University of Durham,
Durham, DH1 3LE }
\end{center}

\vspace*{4cm}
\begin{abstract}
The most general form of the $WWZ$ and $WW\gamma$ interaction
 contains a $CP$-violating term
which has the same threshold behaviour as the Standard Model $\ee \to
\ww$ cross section. We calculate the cross section as a function
of the corresponding anomalous coupling, and estimate the bounds which
can be obtained from a measurement of the threshold cross section at
LEP2. We show how the effect of the coupling is most pronounced in the angular
distributions of the final-state fermions.
\end{abstract}

\newpage
One of the most important physics goals of the LEP2 
$\ee$ collider ($\sqrt{s} \simeq 160 - 190\; \GeV$)
is to measure or constrain anomalous couplings, in particular
the trilinear couplings  of the $W$, $Z$ and photon.
In the SU(2)$\times$U(1)
 Standard Model (SM) these couplings are completely 
specified, and so any deviations from the SM values would signal new physics.
A very detailed and up-to-date
 study of how the various couplings can be measured
at LEP2 can be found in Ref.~\cite{YELLOWANOM}. 
The idea is to use the 
angular distributions of the decay products in $\ee \to \ww
\to 4f$ production, which are sensitive to the form of 
the $WW\gamma$ and $WWZ$ vertices.

Another important measurement at LEP2 is the determination of
the $W$ boson mass ($M_W$) \cite{YELLOWMW}. Two methods have been discussed:
the `direct reconstruction' of the mass from the decay products,
and the measurement of the total $\ww$ cross section close to the 
threshold at $\sqrt{s} = 2 M_W$, which is sensitive to $M_W$.
Here one determines the $W$ mass from equating
the SM prediction and the experimental measurement, with $M_W$ as
a free parameter: $ \sigma_{SM}(M_W \pm \Delta M_W) = 
\sigma_{exp} \pm \Delta \sigma_{exp}$. 
It has been shown \cite{YELLOWMW,WJS} that a collision
energy of $161\; \GeV$ is optimal in this respect. 
An apparent  drawback to this method is that it assumes that the theoretical 
cross section is correctly given by the SM. However it is very difficult 
to imagine {\it any} type of new physics corrections which would
significantly alter the threshold cross section.
The key point is that at threshold the total cross section is dominated
by the $t$-channel neutrino exchange diagram (see Fig.~1 below).
Contributions from $s$-channel photon and $Z$ exchange
are suppressed by a relative factor $\beta^2 = 1 - 4M_W^2/s \ll 1$.
This means that contributions from new physics processes
such as $\gamma^*,Z^*\to X \overline{X}$, or anomalous 
contributions to $\gamma^*, Z^* 
\to \ww$ are heavily suppressed. Loop corrections involving new
particles are also either very small, or else part of the 
renormalized $W$ propagator and therefore  
included in the definition of $M_W$ itself.

There is one important exception to this rule. The
most general effective Lagrangian for the $WWV$ vertex ($V=Z,\gamma$)
in $\ee\to\ww$ production contains  a total of seven
distinct couplings \cite{HAGIWARA}, see Eq.~(\ref{eq:vert}) below, 
 each with different 
properties under discrete $C$, $P$ and $T$ transformations.
All but one of these couplings are suppressed by  a factor $\beta^2$
at threshold, including of course the SM couplings. The exception
is  the `$f_6^V$' (in the notation of Ref.~\cite{HAGIWARA})
 $CP$-violating coupling, which has the {\it same} threshold behaviour
 as the leading SM $\nu$-exchange contribution.\footnote{There is 
 a straightforward angular momentum argument for this, 
 see for example Ref.~\cite{HAGIWARA}.}
The validity of the threshold cross section method for 
determining $M_W$ must therefore rely on the assumption 
that this coupling is either 
zero or very small. 

In this letter we address the following question: if we assume that
there is indeed a non-zero $f_6^V$ $CP$-violating $WWV$ coupling,
what information on it  can be obtained from
 a measurement of the threshold  $\ww$ cross section at LEP2?
To do this we calculate the cross section as a function
of $f_6^V$ and use the expected experimental precision
to obtain an estimate for the uncertainty in its determination.
Of course now one has to {\it assume}  a value for $M_W$,
for example from the direct measurements at $p \bar p$ colliders.
We also study the effect of the anomalous coupling on several
angular distributions. 

There is an important caveat to this approach. 
Neutron electric dipole moment data already rule out 
an anomalous $CP$-violating $WW\gamma$ coupling 
greater than $O(10^{-3})$ (in units of $e$) \cite{xxx},
unless one allows for fine tuning at this level between 
different contributions. If one further argues
(see for example Ref.~\cite{DERUJULA}) that any extension of the SM
should respect the SU(2)$\times$U(1) gauge symmetry, then 
the same order of magnitude limit should apply also 
to the $WWZ$ coupling, and there would be no observable
effect at LEP2. We believe, however, that there is 
no substitute for a {\it direct} measurement. If one
{\it did} discover a $CP$-violating $WWV$ coupling 
at LEP2, the theoretical implications would be immense.
For this reason, we consider  both non-zero $\fg$ and $\fz$
couplings in our study.

The  most general $WWV$ vertex which exhausts all possible
Lorentz structure for $\ww$ production has the form \cite{HAGIWARA}:
\begin{eqnarray}
g_{WWV}^{-1}\; \Gamma_V^{\alpha\beta\mu}(k_-,k_+,P)&=&
        f^V_1(k_- - k_+)^\mu g^{\alpha\beta}
   - {f^V_2 \over M^2_W} (k_- - k_+)^\mu P^\alpha P^\beta \nonumber\\
& &  +f^V_3(P^\alpha g^{\mu\beta} - P^\beta g^{\mu\alpha}) 
    +if^V_4(P^\alpha g^{\mu\beta}+P^\beta g^{\mu\alpha}) \nonumber\\
& & +if^V_5 {\epsilon}^{\alpha\beta\mu\rho} (k_- - k_+)_{\rho} 
    - f_6^V \epsilon^{\mu\alpha\beta\rho} P_{\rho} \nonumber\\
& & -{f^V_7 \over M^2_W} (k_- - k_+)^\mu{\epsilon}^{\alpha\beta\rho\sigma} 
P_\rho (k_- - k_+)_\sigma \; ,
\label{eq:vert}
\end{eqnarray}
where
\begin{equation}
g_{WW\gamma}=-e  \quad  , \quad
g_{WWZ}= -e \cot\theta_W  \; .
\end{equation}
Here $k_\pm$ are the four-momenta of the outgoing $W^\pm$ and $P$
is the incoming four-momentum of the neutral boson $V=\gamma, Z$.
The couplings $f^V_{1,2,3}$ in Eq.~(\ref{eq:vert}) are 
$C$ and $P$ conserving, while the remainder
are $C$ and/or $P$ violating. In particular, the $f^V_6$ coupling 
is $C$ conserving, but $P$ violating. More importantly in the present
context, it is the only coupling that gives a leading
threshold behaviour. In what follows we set all the other $f_i^V$
to their SM values, i.e.
\begin{equation}
f^V_1 = \half f^V_3 = 1\; , \quad f^V_2 =f^V_4 =f^V_5 =f^V_7 = 0\; .
\end{equation}

At LEP2 energies, i.e. far above the $Z$ pole, the $\fg$ and $\fz$
contributions have a comparable effect on the total cross section and
angular distributions. From a measurement with modest luminosity at 
a single threshold energy,
it will therefore  be very difficult to distinguish separate $\fg$ and
$\fz$ contributions. In our numerical studies we therefore set
$\fg = \fz \equiv f_6$.
Using, for example, the spinor techniques of Ref.~\cite{SPINOR} it is
straightforward to compute the $\ee\to\ww\to 4f$ scattering
 amplitude\footnote{Only diagrams with two resonant $W$ propagators
 are included.}
including the anomalous $f_6$ contribution.
The general form is
\begin{equation}
\cM_{WW} = \cM_\nu + \cM_V + \cM_{f_6} \; ,
\label{eq:amps}
\end{equation}
where the contributions on the right-hand side correspond respectively
to $t$-channel neutrino exchange, Standard Model
$s$-channel $\gamma$ and $Z$ exchange, and the anomalous contribution.
For positive helicity initial-state electrons the first of these is
absent.

We begin by considering the total $\ee\to\ww$ cross section in the
zero width (stable $W$) limit, at leading order.
Unless otherwise stated, the numerical
calculations described below use the same set of input parameters
as the study of Ref.~\cite{WJS}. Where $M_W$ is needed as an input
parameter, the recent world average value \cite{WORLDMW} from
direct measurements at the $p \bar p $ colliders,
\begin{equation}
M_W = 80.33 \pm 0.15\; \GeV      \; ,
\end{equation}
is used.

Fig.~1 shows  the contributions to the cross section
corresponding to the decomposition of Eq.~(\ref{eq:amps}), with 
$f_6 = 1$ for illustration,
as a function of the collider energy $\sqrt{s}$. Note that 
there is no interference between the
SM and $f_6$ amplitudes in the total cross section. 
Just above threshold the dependence of the cross section
on the $W$ velocity, $\beta = \sqrt{1 - 4 M_W^2/s}$, can be parametrized
as
\begin{equation}
\sigma = A \beta + B \beta^3 + \ldots \; .
\end{equation}
The threshold behaviour discussed above is clearly evident: the
anomalous contribution is $O(\beta)$, as for the
$\nu$-exchange contribution and in contrast to the $O(\beta^3)$
behaviour of the SM $s$-channel contributions. For $f_6 = 1$, the ratio
of $\Delta\sigma_{f_6}$ to $\sigma_{SM}$ is approximately $0.43$ near
 threshold, in agreement with Ref.~\cite{HAGIWARA2}.\footnote{The
calculation of Ref.~\cite{HAGIWARA2} assumed $\fg = 0$, in which 
case the ratio is $0.29$.}

In practice the stable $W$  approximation is inadequate in the threshold
region. There are important contributions from finite width effects,
initial-state radiation, Coulomb corrections etc. \cite{YELLOWMW},
all of which smear out the sharp rise from zero of the cross sections in
Fig.~1. These effects are included in our calculation, exactly
as described in Ref.~\cite{WJS}. 
 
Fig.~2 shows the total $WW$ cross section in the 
 threshold region as a function of $\sqrt{s}$,
for fixed $M_W = 80.33\; \GeV$ and  different $f_6$ values.
Since there is no interference between the SM and $f_6$ amplitudes,
the cross section depends quadratically on $f_6$ at a given energy.
The curves are reminiscent of the behaviour of the threshold
cross section on $M_W$ (see for example Fig.~4 of Ref.~\cite{WJS}),
with one important difference: the sensitivity of the cross section
to $f_6$ (as parametrized by the ratio of $\delta\sigma/\sigma$ 
to $\delta f_6/f_6$) is approximately independent of $\sqrt{s}$ in
the threshold region. This contrasts with the corresponding
sensitivity to $M_W$, which is maximal roughly $500\; \MeV$
above the nominal threshold at $\sqrt{s} = 2 M_W$ \cite{YELLOWMW,WJS}.
It is for this reason that the `threshold running' of LEP2 will
take place at the single collision 
 energy $\sqrt{s} = 161\; \GeV$. In the remainder of this paper, therefore,
 we restrict our attention to this value.
 
Fig.~3 shows $\sigma_{WW}(161\; \GeV)$ as a function of  $f_6$,
for $M_W = 80.33 \pm 0.15\; \GeV$,  the current world average.
The expected quadratic behaviour is clearly evident. We can use this figure
to estimate an approximate experimental error on $f_6$ from a total 
cross section measurement. In Ref.~\cite{YELLOWMW}  it was estimated
that for 4 experiments each with $50$~pb$^{-1}$ total luminosity
the error on the $W$ mass would be $\delta M_W = \pm 108\; \MeV$.
This corresponds to $\delta\sigma/\sigma \approx 1/16$, indicated
by the horizontal band in Fig.~3 centred on the SM prediction
evaluated at the world average $M_W$. Fixing $M_W$ at this value
gives $\delta f_6 = \pm 0.4$ for the same cross section uncertainty.
If the current $\pm 150\; \MeV$ error on $M_W$ is taken into account,
this increases to $\delta f_6 = \pm 0.6$.

Information will also be available on the distribution of the $\ww$
decay products, for example the angular distributions of the final-state
leptons and jets. These can provide important additional constraints,
particularly since in such distributions the interference
between the SM and $CP$-violating amplitudes leads to a  {\it linear }
dependence on $f_6$. Given the relatively small statistics, the 
difficulty in reconstructing the quark momenta from the observed jets,
and the presence of at least one energetic neutrino in more than half
the events, only rather simple distributions are likely
to be accessible in practice.
Nevertheless, these can still be rather sensitive to a possible
 $CP$-violating contribution. As an example, 
Fig.~4(a) shows the (laboratory frame) 
polar angle distribution of the charged lepton
$l^-=e^-,\mu^-,\tau^-$ from $W^- \to l^-\bar{\nu}_l$ decay.
The curves show the SM and anomalous contributions for $f_6 = 1$.
Notice the large negative contribution at small angles from the 
interference between the SM and anomalous amplitudes, which has the effect
of producing a pronounced dip  in the overall distribution.
An important signature of the $CP$-violating properties of the anomalous
interaction is the fact that the distribution in Fig.~4(a) is {\it not}
symmetric under the interchange: $l^- \leftrightarrow l^+$,
$ \cos\theta \leftrightarrow -\cos\theta$. To illustrate this, we define
the asymmetry 
\begin{equation}
A = {d\sigma/d\cos\theta(l^-,\cos\theta) 
            - d\sigma/d\cos\theta(l^+,-\cos\theta)
\over d\sigma/d\cos\theta(l^-,\cos\theta)
            + d\sigma/d\cos\theta(l^+,-\cos\theta) }\; ,
\label{eq:asy}
\end{equation}
which vanishes in the SM.  Fig.~4(b) shows $A$ as a function of 
$\cos\theta$ for $f_6 = 1$.            
The measurement of such distributions 
 can be used to improve the precision of the 
$f_6$ determination, although obtaining a realistic, quantitative
 estimate will require
a detailed detector simulation beyond the scope of the present work.

As a final example, Fig.~5 shows the distribution in the angle between the 
normals of the two planes containing the fermions from each $W$ decay,
i.e.
\begin{equation}
\cos\varphi = { (\bm{p}_1 \times \bm{p}_2) \cdot
 (\bm{p}_3 \times \bm{p}_4) \over
\vert\bm{p}_1\vert\vert\bm{p}_2\vert\vert\bm{p}_3\vert\vert\bm{p}_4\vert}
\; ,
\label{cross}
\end{equation}
where $\bm{p}_1$ ($\bm{p}_2$) labels the momentum of the 
fermion (antifermion) from the $W^-$, and $\bm{p}_3$ ($\bm{p}_4$) labels the 
momentum of the antifermion (fermion) from the $W^+$.
 Since the relative orientation with respect to the 
incoming leptons has been integrated out, there is no
 interference between the SM and $CP$-violating amplitudes.
The individual contributions are evidently very different, however,
with the SM ($CP$-violating) contribution preferring the planes
to be aligned (anti-aligned). Note that the definition in Eq.~(\ref{cross})
assumes an ideal situation where the fermions and antifermions
can be identified in the $W$ decay products. 
This is could in principle be achieved  for final state quarks jets 
by a jet-charge analysis or by requiring a charm quark jet,
but in practice the limited statistics are likely to make this difficult.

In conclusion, we have studied the effect on the threshold $\ww$
cross section at LEP2 of a possible $CP$-violating 
$WWV$ interaction, parametrized by couplings $\fg$ and $\fz$.
 Although there 
is indirect evidence that such couplings are very small, we believe that 
a direct search is important, in view of the implications for
the $W$ mass measurement from the threshold cross section. We
have estimated the likely precision on the $f_6^V$ measurements,
and studied several angular distributions which will provide
further information. 

\section*{Acknowledgements}

This work  was supported in part  by the EU under the Human 
Capital and Mobility Network Program CHRX--CT93--0319.

\newpage
\section*{Figure Captions}
\begin{itemize}
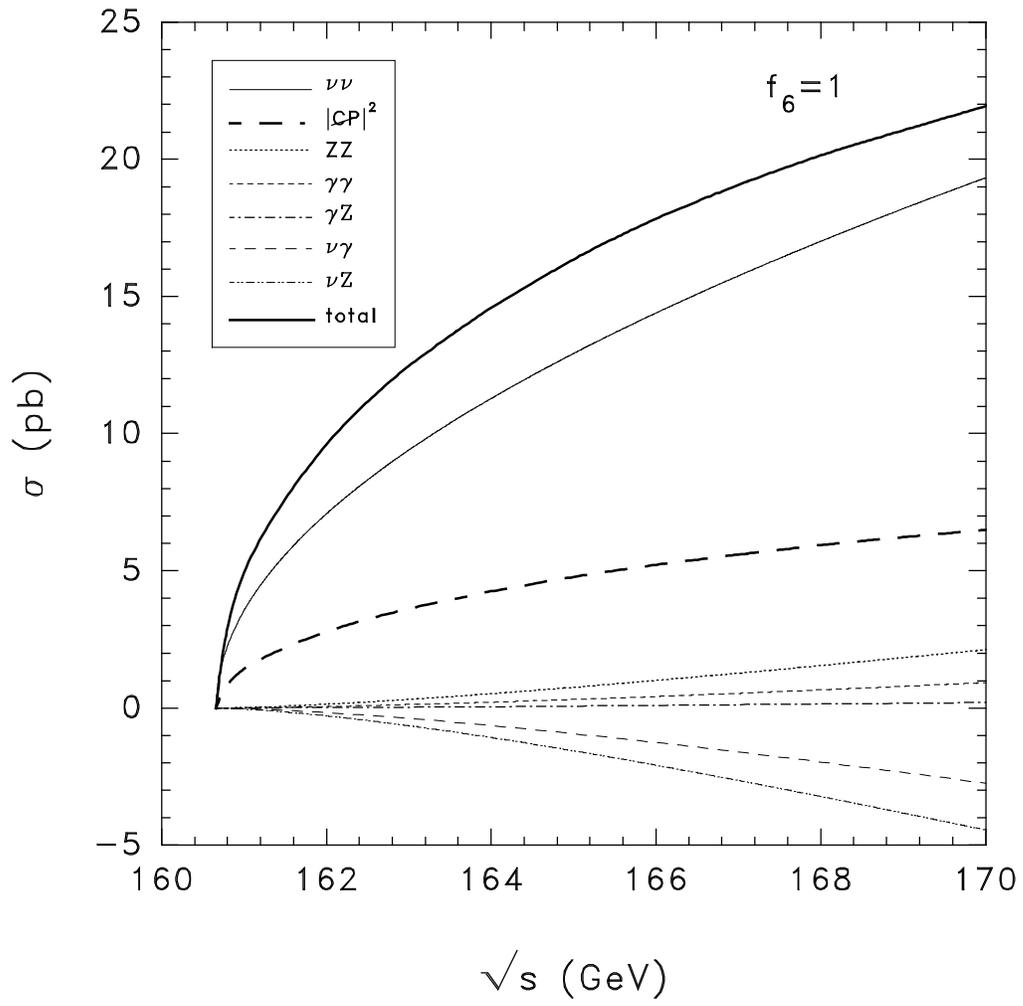
\item[Fig.~1]
Decomposition of the (on-shell) Born $\ee\to\ww$  cross section
 into its various SM components, together with 
 the $f_6=1$ $CP$-violating contribution, as a function of $\sqrt{s}$.

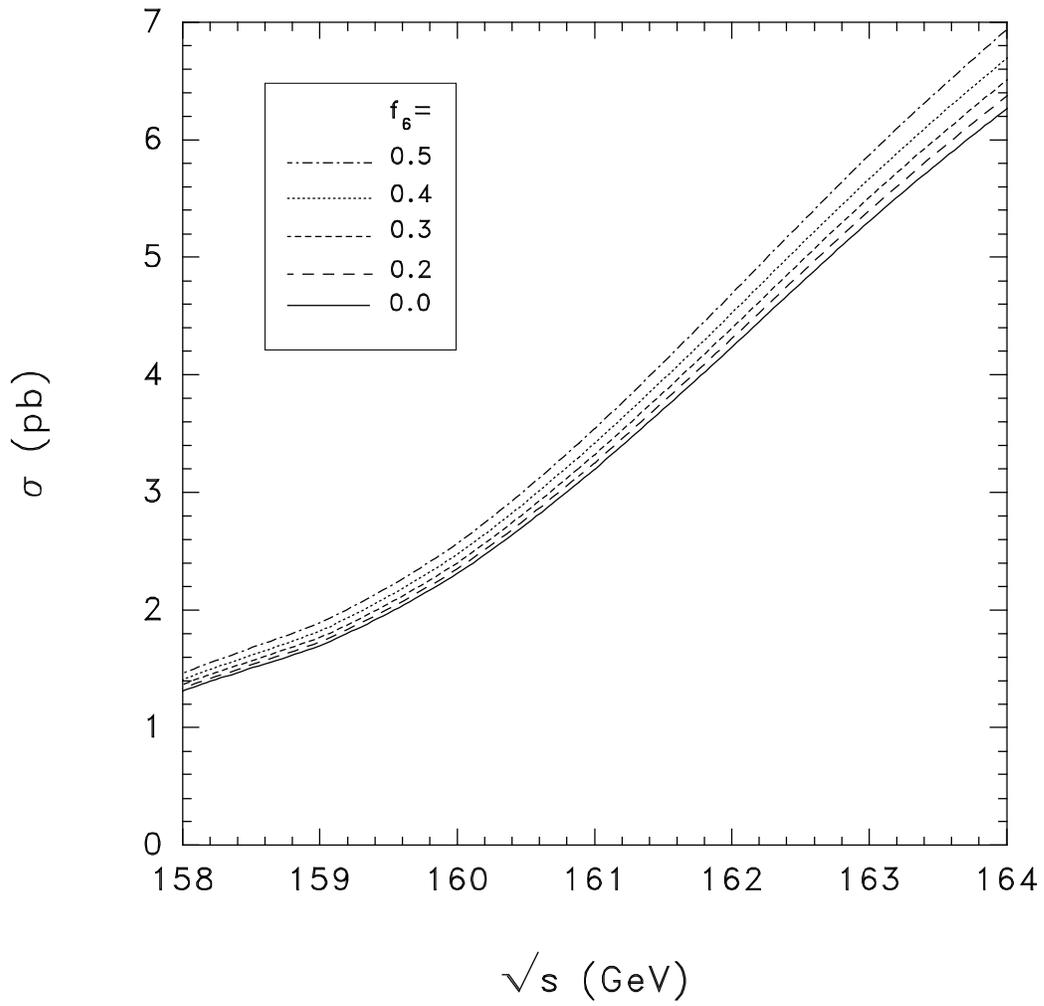
\item[Fig.~2]
The total (off-shell) $\ee\to\ww$ cross section, including ISR
and Coulomb corrections,  for 
various values of the $f_6$ coupling, as a  function of $\sqrt{s}$.

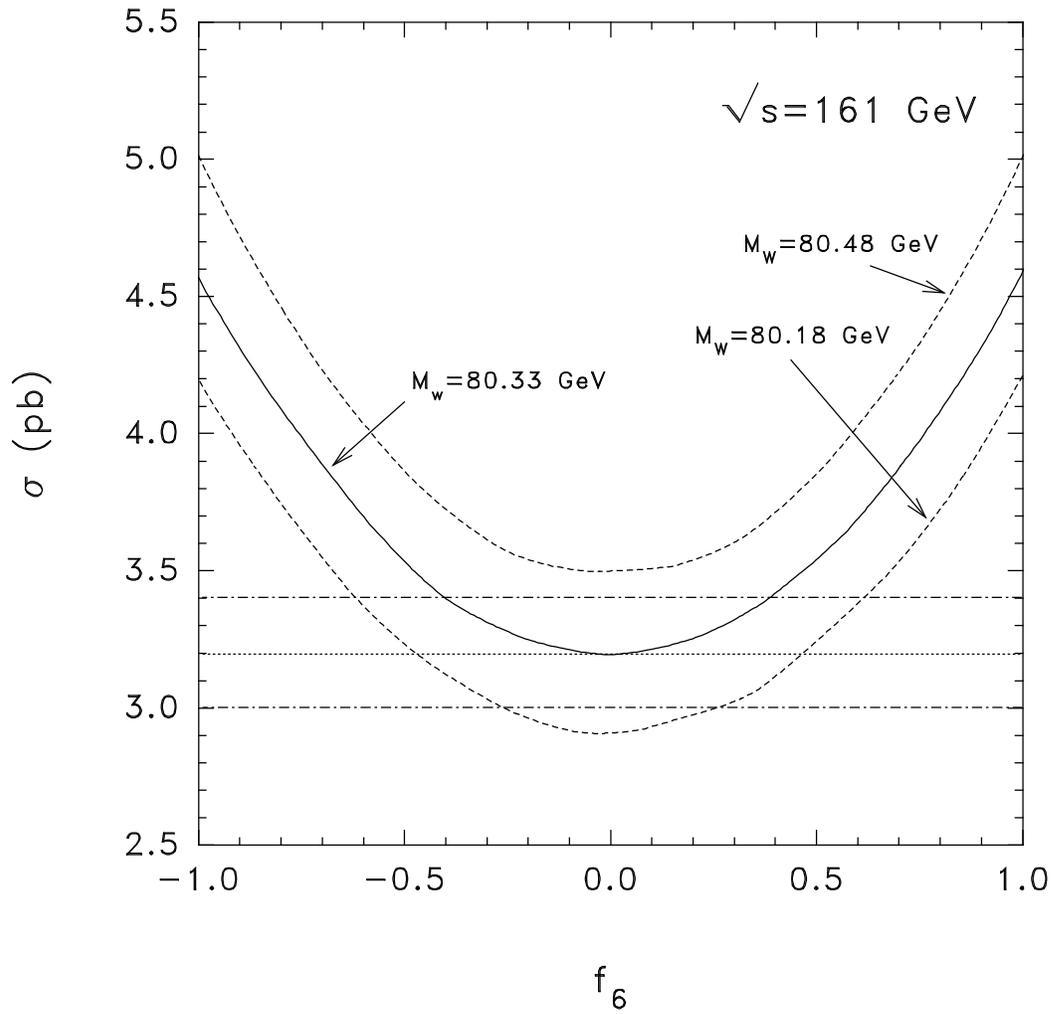
\item[Fig.~3]
The total  $\ee\to\ww$ cross section at 
$\sqrt{s} = 161\; \GeV$, as a function of $f_6$. The solid line corresponds
to $M_W=80.33\; \GeV$, while the band defined by the short-dashed
lines represents the experimental uncertainty of $\pm 0.15\; \GeV$. 
The horizontal band indicates a possible experimental measurement,
as discussed in the text.

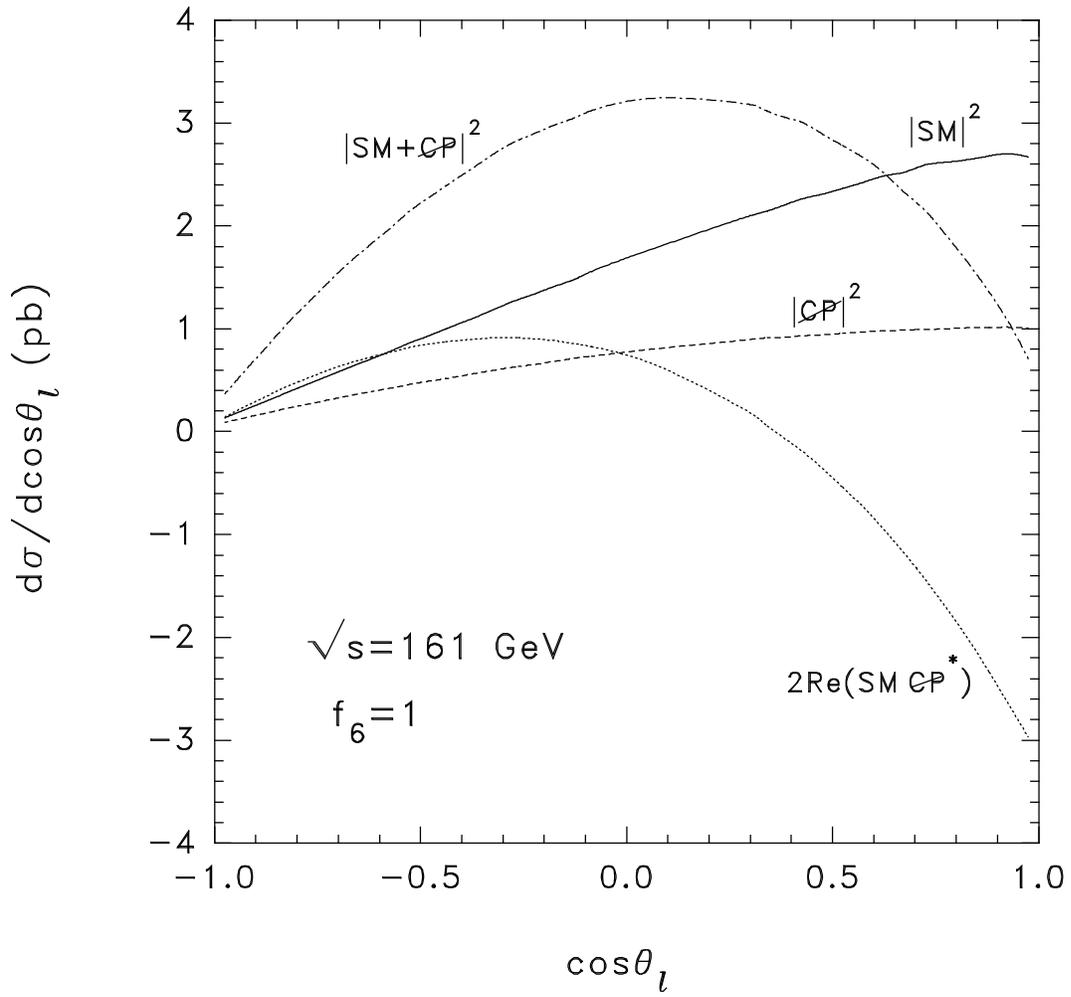
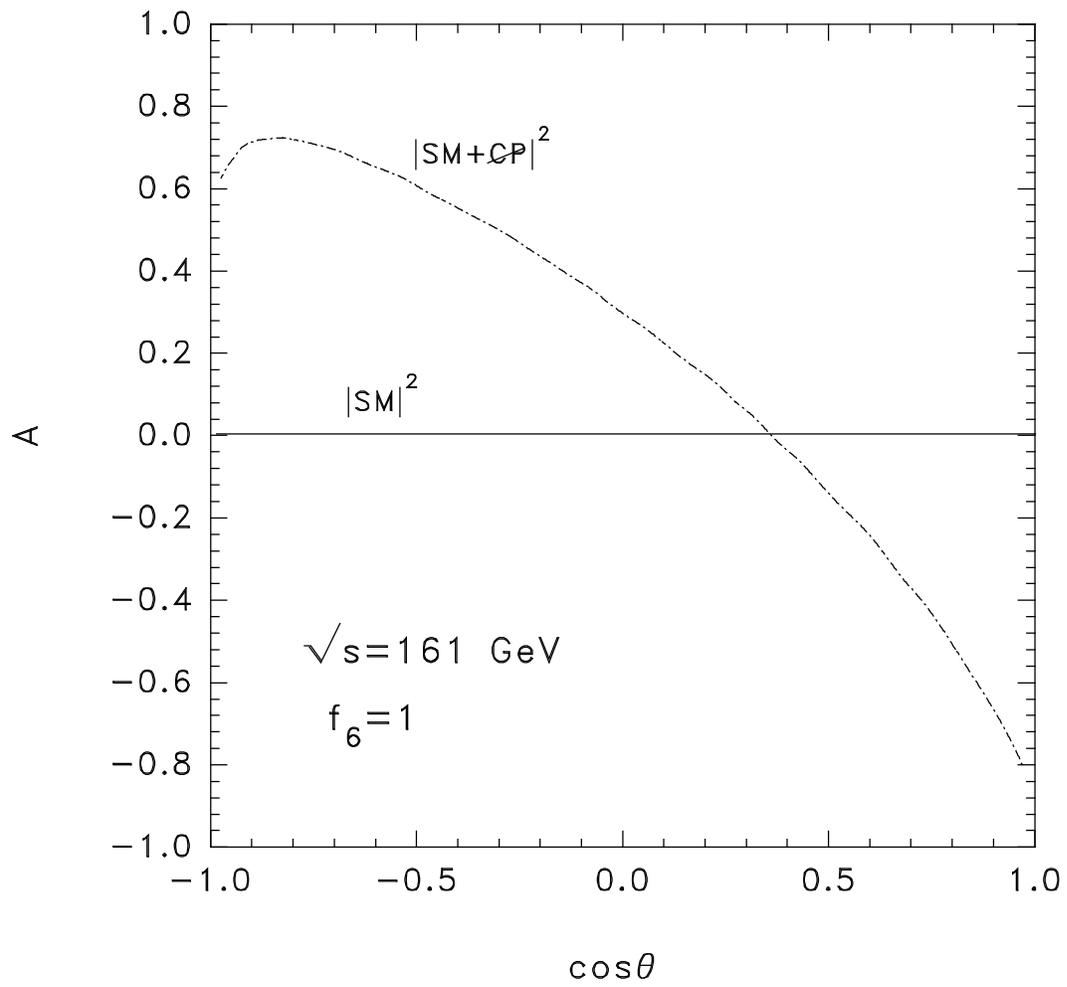
\item[Fig.~4]
Distributions  in (a) the lepton ($l^-$)  polar angle and (b)
the $l^\pm$ forward-backward asymmetry  defined in Eq.~(\ref{eq:asy}), at 
$\sqrt{s} = 161\; \GeV$, for  $f_6=1$.

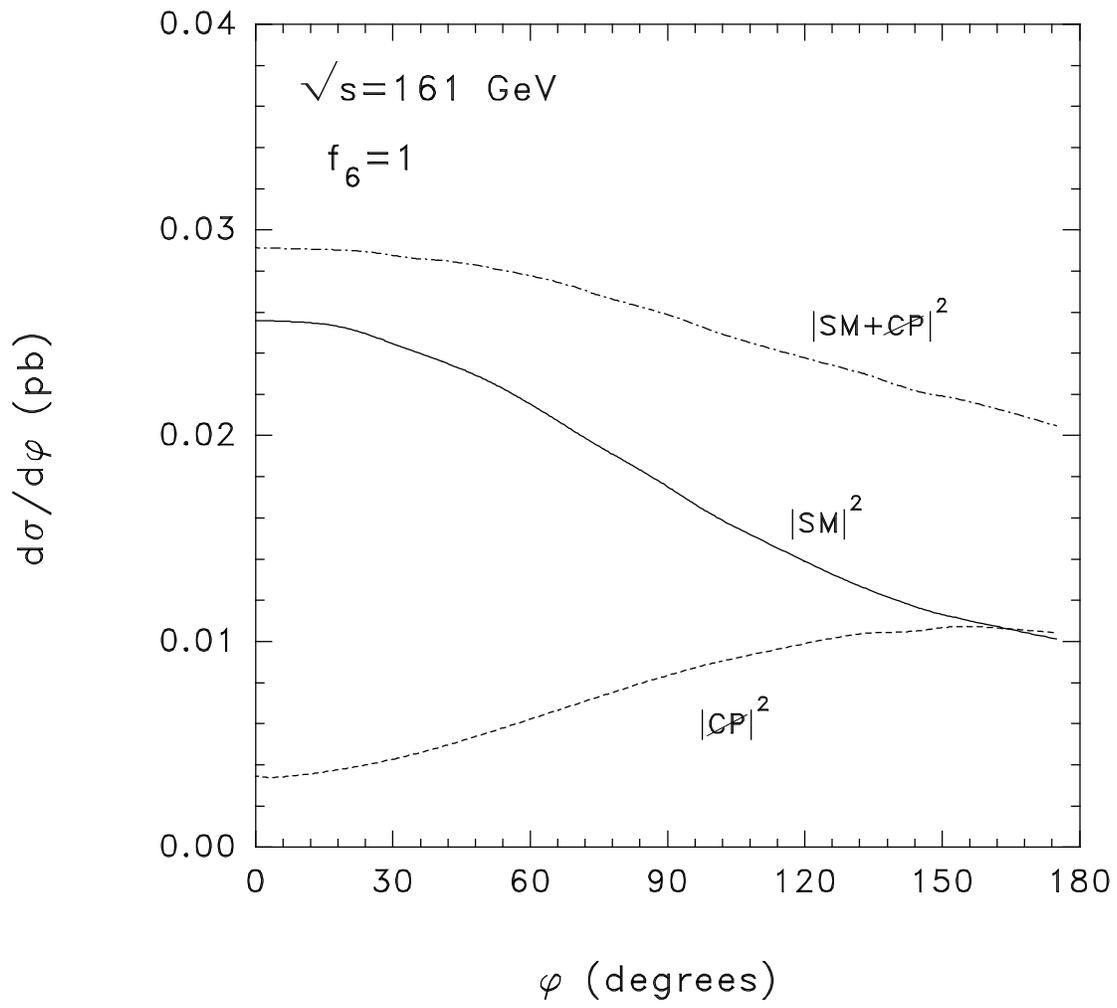
\item[Fig.~5]
Distribution in the angle between the normals to the planes
of the $W^\pm$ decay products at $\sqrt{s} = 161\; \GeV$, for $f_6=1$.

\end{itemize}
\end{document}